\documentclass[letter,longauth]{aa} 

\usepackage{graphicx}
\usepackage{txfonts}
\usepackage{xcolor}
\usepackage{natbib}

\usepackage[colorlinks=true,linkcolor=blue,citecolor=magenta,pdfborder={0 0 0}]{hyperref}
\usepackage{amsmath}

\newcommand{\kms}{\mbox{km\,s$^{-1}$}}
\newcommand{\objone}{\mbox{J$0925$}}

\newcommand{\lya}{\relax \ifmmode {\mbox Ly}\alpha\else Ly$\alpha$\fi}
\newcommand{\ha}{\relax \ifmmode {\mbox H}\alpha\else H$\alpha$\fi}
\newcommand{\hg}{\relax \ifmmode {\mbox H}\gamma\else H$\gamma$\fi}
\newcommand{\hd}{\relax \ifmmode {\mbox H}\delta\else H$\delta$\fi}
\newcommand{\hb}{\relax \ifmmode {\mbox H}\beta\else H$\beta$\fi}
\newcommand{\sii}{\relax \ifmmode {\mbox S\,{\scshape ii}}\else S\,{\scshape ii}\fi}
\newcommand{\nii}{\relax \ifmmode {\mbox N\,{\scshape ii}}\else N\,{\scshape ii}\fi}
\newcommand{\neiii}{\relax \ifmmode {\mbox Ne\,{\scshape iii}}\else Ne\,{\scshape iii}\fi}
\newcommand{\oii}{\relax \ifmmode {\mbox O\,{\scshape ii}}\else O\,{\scshape ii}\fi}
\newcommand{\oi}{\relax \ifmmode {\mbox O\,{\scshape i}}\else O\,{\scshape i}\fi}
\newcommand{\oiii}{\relax \ifmmode {\mbox O\,{\scshape iii}}\else O\,{\scshape iii}\fi}
\newcommand{\hii}{\relax \ifmmode {\mbox H\,{\scshape ii}}\else H\,{\scshape ii}\fi}
\newcommand{\hi}{\relax \ifmmode {\mbox H\,{\scshape ii}}\else H\,{\scshape i}\fi}
\graphicspath{{./}{figures/}}

\begin{document}

   \title{Ubiquitous broad-line emission and the relation between ionized gas outflows and Lyman continuum escape in Green Pea galaxies}
   \titlerunning{Outflows and turbulence in Lyman continuum emitters}
\authorrunning{Amor\'in et al.}
\author{R.~O.~Amor\'in\inst{\ref{inst1},\ref{inst2}}
\and
    M. Rodr\'iguez-Henr\'iquez\inst{\ref{inst3},\ref{inst2}}
    \and    V.~Fern\'andez\inst{\ref{inst4},\ref{inst5}}
    \and
    J.~M.~Vílchez\inst{\ref{inst6}}
    \and
    R.~Marques-Chaves\inst{\ref{inst7}}
    \and
    D.~Schaerer\inst{\ref{inst7}}
    \and
    Y.~I.~Izotov\inst{\ref{inst8}}
    \and
    V. Firpo\inst{\ref{inst9}}
    \and
    N. Guseva\inst{\ref{inst8}}
    \and
     A.~E.~Jaskot\inst{\ref{inst10}}
    \and
    L.~Komarova\inst{\ref{inst4}}
    \and
    D. Muñoz-Vergara\inst{\ref{inst2}}
    \and
     M.~S.~Oey\inst{\ref{inst4}}
    \and
    O. Bait\inst{\ref{inst7}}
    \and
    C.~Carr\inst{\ref{inst11},\ref{inst12}}
    \and
    J.~Chisholm\inst{\ref{inst13}}
    \and
    H. Ferguson\inst{\ref{inst14}}
    \and
    S.~R.~Flury\inst{\ref{inst10}}
    \and
    M.~Giavalisco\inst{\ref{inst10}}
    \and
    M.~J.~Hayes\inst{\ref{inst15}}
    \and
    A.~Henry\inst{\ref{inst16},\ref{inst14}}
    \and
    Z.~Ji\inst{\ref{inst17}}
    \and
    W.~King\inst{\ref{inst13}}
    \and
    F.~Leclercq\inst{\ref{inst13}}
    \and
    G.~\"Ostlin\inst{\ref{inst15}}
    \and
     L.~Pentericci\inst{\ref{inst19}}
    \and
    A.~Saldana-Lopez\inst{\ref{inst15}}
    \and
    T.~X. Thuan\inst{\ref{inst20}}
    \and
    M.~Trebitsch\inst{\ref{inst21}}
    \and
   B.~Wang\inst{\ref{inst22},\ref{inst23},\ref{inst24}}
   \and
    G.~Worseck\inst{\ref{inst25}}
    \and
    X.~Xu\inst{\ref{inst26}}
}
    \institute{ARAID Foundation. Centro de Estudios de F\'{\i}sica del Cosmos de Arag\'{o}n (CEFCA), Unidad Asociada al CSIC, Plaza San Juan 1, E--44001 Teruel, Spain\label {inst1}
\and
    Departamento de Astronom\'ia, Universidad de La Serena, Avda. Juan Cisternas 1200, La Serena, Chile \label{inst2}
\and
    Gemini Observatory, 670 N. A’ohoku Place, Hilo, Hawai'i, 96720, USA \label{inst3}
\and
    University of Michigan, Department of Astronomy, 323 West Hall, 1085 S. University Ave, Ann Arbor, MI 48109, USA \label{inst4}
\and
    Instituto de Investigaci\'on Multidisciplinar de Investigación y Posgrado, Universidad de La Serena, Ra\'ul Bitr\'an 1305, La Serena, Chile \label{inst5}
\and
    Instituto de Astrofísica de Andalucía, CSIC, Apartado de correos 3004, E-18080 Granada, Spain \label{inst6}
\and
    Department of Astronomy, University of Geneva, 51 Chemin Pegasi, 1290 Versoix, Switzerland \label{inst7}
\and
    Bogolyubov Institute for Theoretical Physics, National Academy of Sciences of Ukraine, 14-b Metrolohichna str., Kyiv, 03143, Ukraine
    \label{inst8}
\and
    Gemini Observatory/NSF’s NOIRLab, Casilla 603, La Serena, Chile \label{inst9}
\and
    Department of Astronomy, University of Massachusetts, Amherst, MA 01003, USA \label{inst10}
\and
    Center for Cosmology and Computational Astrophysics, Institute for Advanced Study in Physics, Zhejiang University, Hangzhou 310058, China\label{inst11}
\and
    Institute of Astronomy, School of Physics, Zhejiang University, Hangzhou 310058,  China\label{inst12}
\and
    Department of Astronomy, The University of Texas at Austin, 2515 Speedway, Stop C1400, Austin, TX 78712-1205, USA \label{inst13}
\and
    Space Telescope Science Institute, 3700 San Martin Drive, Baltimore, MD 21218, USA \label{inst14}
\and
    Department of Astronomy, Oskar Klein Centre; Stockholm University; SE-106 91 Stockholm, Sweden \label{inst15}
\and
    Center for Astrophysical Sciences, Department of Physics \& Astronomy, Johns Hopkins University, Baltimore, MD 21218, USA \label{inst16}
\and
    Steward Observatory, University of Arizona, 933 N. Cherry Avenue, Tucson, AZ 85721, USA \label{inst17}
\and
    INAF - Osservatorio Astronomico di Roma, Via di Frascati 33, 00078, Monte Porzio Catone, Italy \label{inst19}
\and
    Astronomy Department, University of Virginia, P.O. Box 400325, Charlottesville, VA 22904-4325, USA \label{inst20}
\and
    Kapteyn Astronomical Institute, University of Groningen, P.O. Box 800, 9700 AV Groningen, The Netherlands \label{inst21}
\and
    Department of Astronomy \& Astrophysics, The Pennsylvania State University, University Park, PA 16802, USA\label{inst22}
\and
    Institute for Computational \& Data Sciences, The Pennsylvania State University, University Park, PA 16802, USA\label{inst23}
\and
    Institute for Gravitation and the Cosmos, The Pennsylvania State University, University Park, PA 16802, USA\label{inst24}
\and
    Institut für Physik und Astronomie, Universität Potsdam, Karl-Liebknecht-Str. 24/25, D-14476 Potsdam, Germany\label{inst25}
\and
    Center for Interdisciplinary Exploration and Research in Astrophysics, Northwestern University, Evanston, IL, 60201, USA \label{inst26}
}

   \date{Received ----; accepted ----}


\abstract{We report observational evidence of highly turbulent ionized gas kinematics in a sample of 20 Lyman continuum (LyC) emitters (LCEs) at low redshift ($z\sim$\,0.3).
Detailed Gaussian modeling of optical emission line profiles in high-dispersion spectra consistently shows that both bright recombination and collisionally excited lines can be fitted as one or two narrow components with intrinsic velocity dispersion of $\sigma$\,$\sim$\,40-100\,\kms, in addition to a broader component with $\sigma \sim$\,100-300\,\kms, which contributes up to $\sim$40\% of the total flux and is preferentially blueshifted from the systemic velocity. We interpret the narrow emission as highly ionized gas close to the young massive star clusters and the broader emission as a signpost of unresolved ionized outflows, resulting from massive stars and supernova feedback.
We find a significant correlation between the width of the broad emission and the LyC escape fraction, with strong LCEs exhibiting more complex and broader line profiles than galaxies with weaker or undetected LyC emission.
 We provide new observational evidence supporting predictions from models and simulations; our findings suggest that gas turbulence and outflows resulting from strong radiative and mechanical feedback play a key role in clearing channels through which LyC photons escape from galaxies.
We propose that the detection of blueshifted broad emission in the nebular lines of compact extreme emission-line galaxies can provide a new indirect diagnostic of Lyman photon escape, which could be useful to identify potential LyC leakers in the epoch of reionization with the JWST.}


   \maketitle
%

%
\section{Introduction} \label{sec:intro}

Understanding cosmic reionization requires the identification of the physical mechanisms driving the escape of ionizing photons from galaxies. 
High-resolution hydrodynamical simulations of galaxy formation demonstrate the crucial role of winds from massive stars and the feedback from  supernovae (SNe)  as the primary drivers of turbulence, momentum, and energy into the interstellar medium (ISM) \citep{Nelson2019}. Star formation-driven outflows are thus key ingredients to carve out optically thin channels in the ISM and enable Lyman continuum (LyC) photons to escape from star-forming galaxies  \citep[e.g.,][]{WiseCen2009,KimmCen2014,Kimm2019,Ma2016,Trebitsch2017,Rosdahl2018,Barrow2020,Kakiichi2021}.
However, while the connection between stellar and SNe feedback and LyC escape predicted by theoretical work has been supported by observational studies \citep[e.g.,][]{Heckman2011,Chisholm2017,Micheva2019,Kim2020,Bait2023}, an empirical relation between outflows and the  LyC escape fraction remains unclear \citep{Marques-Chaves2022a, Mainali2022, Naidu2022}.

Probing the nature and properties of outflows in LyC emitters can provide important clues about the spatial scales and timescales required to create the ISM conditions for LyC escape \citep{Zastrow2013}. Radiative feedback and winds from very young massive stars ($<$\,3Myr)  are expected to contribute first, developing localized, subkiloparsec scale outflows that can erode the parent H{\sc i} cloud, clear dust, and create density-bounded channels or holes that permeate LyC radiation \citep{Ferrara2023}. Later, SNe feedback takes over when the LyC production from massive stars declines rapidly, developing kiloparsec-scale outflows that can clear out the neutral gas halo and maximize the LyC escape \citep[e.g.,][]{Naidu2022}. The two modes could eventually operate sequentially \citep[e.g.,][]{Kakiichi2021,Katz2023}.

New insight into the escape of ionizing photons can be obtained from green pea (GP) galaxies \citep{Cardamone2009} at $z$\,$\sim$\,0.1-0.4, which includes the largest numbers of LyC emitters (LCEs) detected at low redshifts \citep[e.g.,][]{Izotov2016a,Izotov2016b,Izotov2018a,Izotov2018b,Flury2022a}. The GPs are compact galaxies ($R_{50}$\,$<$\,1\,kpc), with high equivalent width (EW$_{5007}$\,$\sim$\,200-2000\AA), high specific star formation rate (sSFR$\sim$\,10$^{-8}$-10$^{-7}$ yr$^{-1}$), and low metallicity \citep[$Z/Z_{\odot}\sim$\,0.2][]{Amorin2010,Amorin2012a} and low dust attenuation \citep{Chisholm2022}. While the most extreme GPs host very young starbursts ($\lesssim$\,2-5 Myr) with ionization properties approaching density-bounded conditions \citep[e.g.,][]{Jaskot2013}, in less extreme GPs we find more evolved signatures of Wolf-Rayet (WR) stars and SNe, which produce intense feedback \citep[e.g.,][]{Amorin2012a, Amorin2012b}. 
Signatures of turbulent ionized gas and outflows 
are often identified in emission lines as high-velocity wings and asymmetries  \citep{Amorin2012b,Bosch2019,Hogarth2020,Komarova2021}. Although most LCEs confirmed so far at $z<0.4$ are GPs, a detailed analysis of their ionized gas kinematics using high-resolution spectra is still lacking.

In this {letter}, we report the first observational evidence of strong feedback from the complex ionized gas kinematics seen in galaxies with confirmed LyC emission at $z$\,$\sim$\,0.3.
We investigate the kinematic imprints of strong turbulence and starburst-driven outflows via detailed modeling of nebular emission lines using high-resolution spectra, and we present a correlation between ionized gas kinematics and LyC escape fraction. Finally, we discuss the implications for LyC escape physics and the use of such kinematic imprints as an indication of LyC leakage in compact low-metallicity starbursts with strong Ly$\alpha$ emission at higher redshifts, which often show broad components in their ionized emission lines \citep[e.g.,][]{Matthee2021,Vanzella2022,Mainali2022,Llerena2023}.

\begin{figure*}[ht!]
\centering
\includegraphics[angle=0,width=0.327\textwidth]{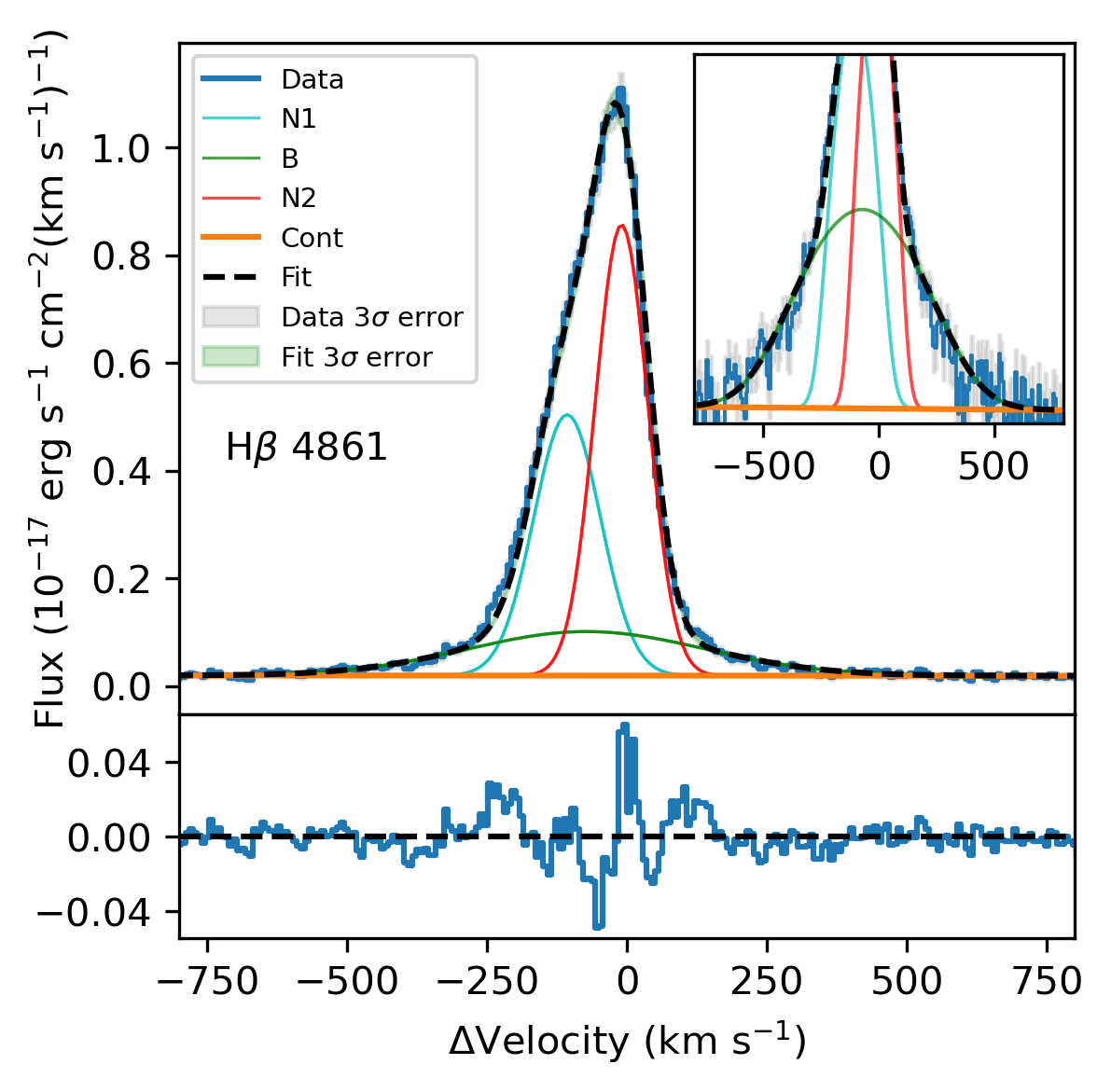}
\includegraphics[angle=0,width=0.318\textwidth]{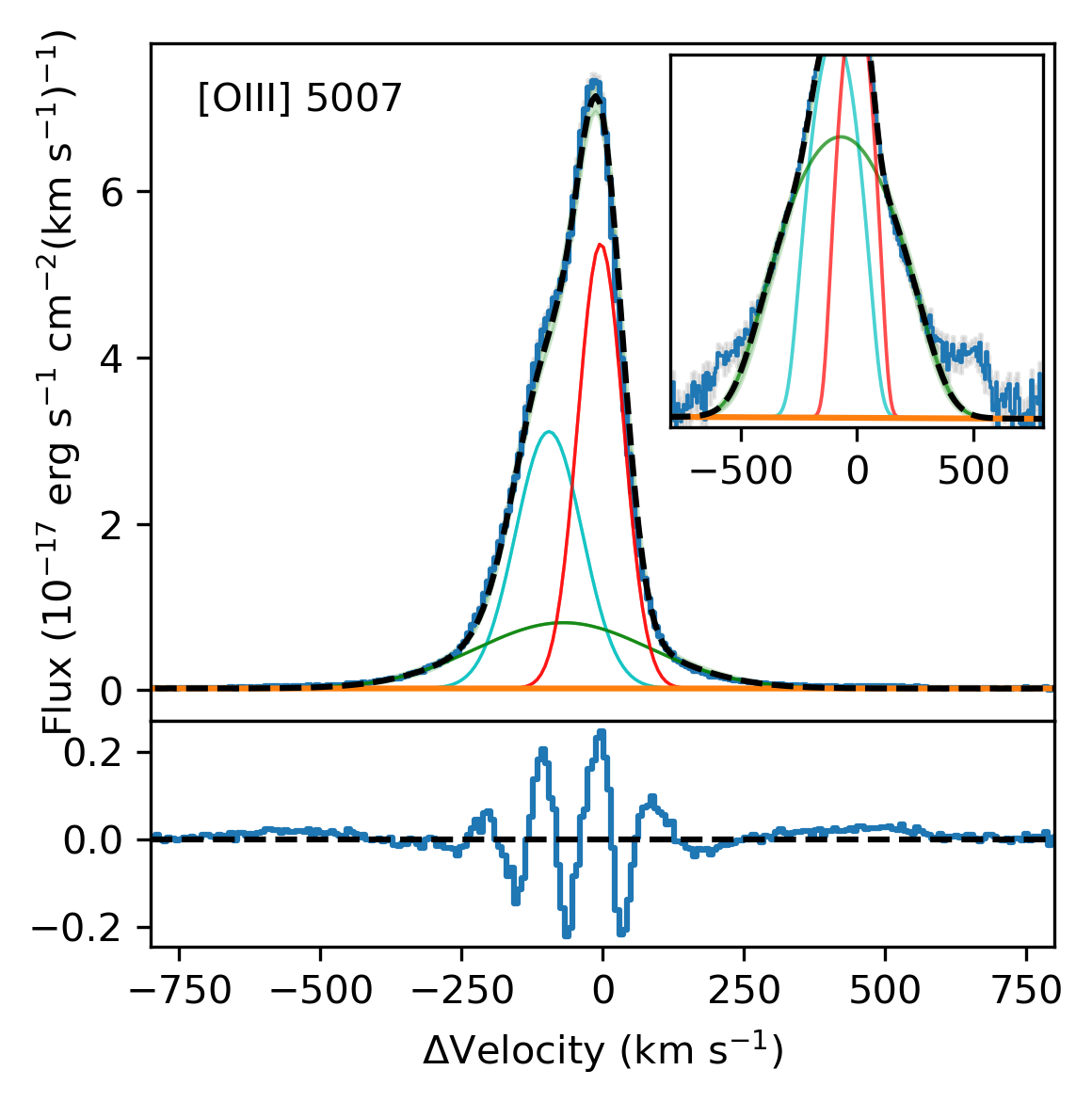}
\includegraphics[angle=0,width=0.32\textwidth]{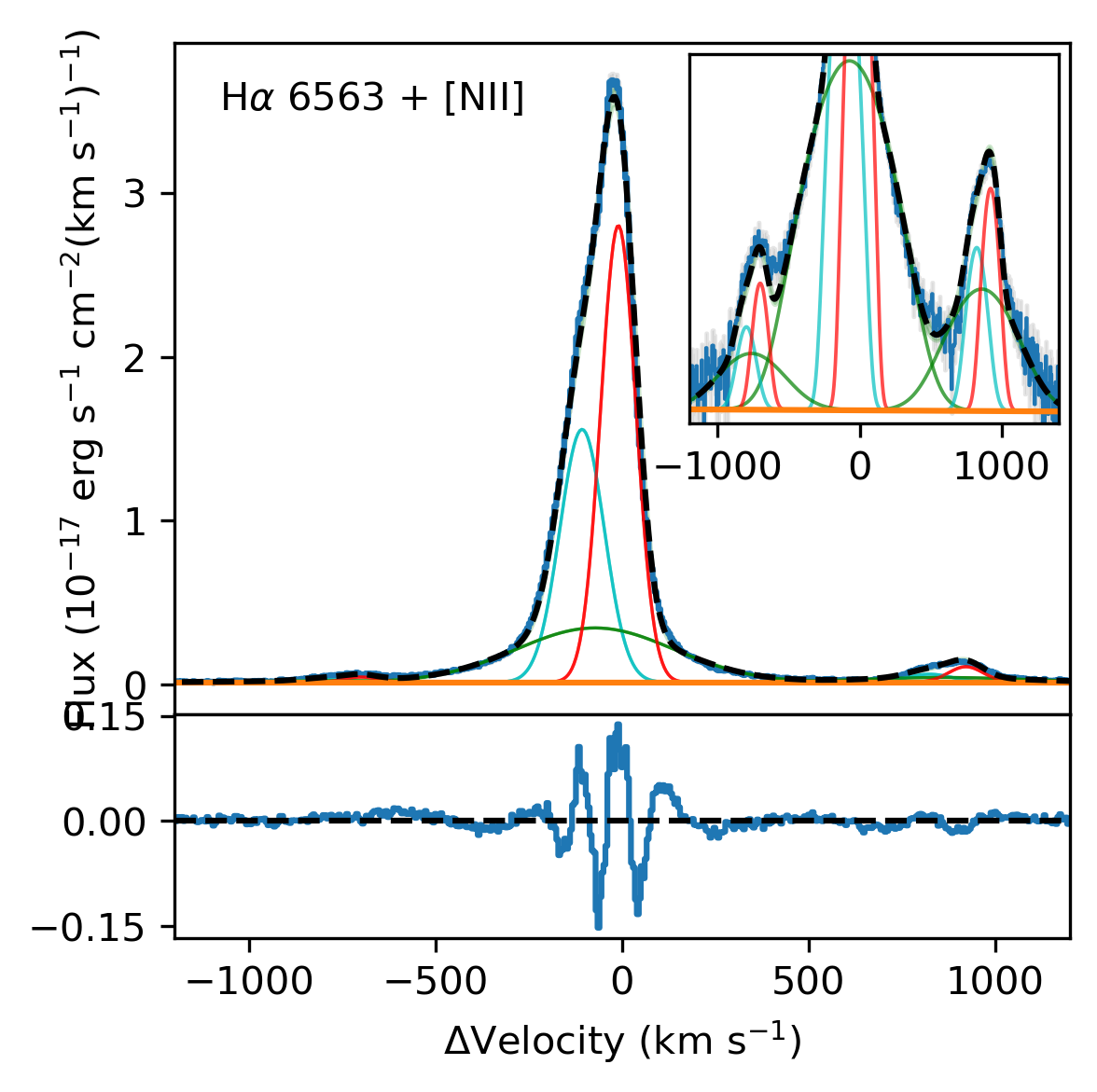}
\caption{Gaussian models (black dashed line) fitted to the observed H$\beta$ (left),  [\oiii]\,$\lambda$\,5007 (center), and H$\alpha +$[N{\sc ii}]\,6748,6584 (right) lines in the WHT/ISIS spectra of the SLCE \mbox{J0925$+$1403}. The bottom panels show residuals in the same flux units. The continuum (orange line) and the two narrow (N1, N2) and broad (B) components are shown in cyan, red, and green, respectively. The inset shows a zoomed-in image  of the line wings.
\label{fig:f1}}
\end{figure*}

\section{Sample and observations} \label{sec:data}

We used a sample of 20 GP galaxies at $z$\,$\sim$\,0.2-0.4 (Fig.~\ref{sample}), including 14 galaxies from the Low-z Lyman Continuum Survey \citep[LzLCS,][]{Flury2022a,Flury2022b,Saldana2022}, 5 galaxies from \citet{Izotov2016a,Izotov2016b,Izotov2018a}, and 1 from \citet{Wang2021} for which high-resolution optical spectra are available. We used the UV spectral properties and images from HST/COS presented in \citet{Flury2022a,Flury2022b}. Following their definitions, 6 galaxies are classified as {\it Strong LCE} (SLCE), 11 as  {\it Weak LCE} (WLCE), and 3 as {\it Non-LCE} (NLCE), according to the significance and the signal-to-noise ratio (S/N) of the LyC detection.

For a subsample of seven galaxies, we used observations obtained with the Intermediate Dispersion Spectrograph and Imaging System (ISIS) on the 4.2 m William Herschel Telescope (WHT, Program P27, PI: R. Amor\'in), following the instrumental configuration presented by \citet{Hogarth2020}.

In short, the R1200B and R1200R gratings were used for the light split by the D6100 dichroic into the blue and red arms of the spectrograph, respectively;  each arm was centered around the observed wavelengths of \hb\ and \ha.
We used a long slit 0\farcs9 in width, oriented at the parallactic angle.
Observing nights were  non-photometric and with an average seeing of 1\arcsec.
The average spectral dispersion and full width at half maximuum (FWHM), as measured on sky lines and arcs of the blue (red) arm, were 0.23 (0.26) \AA\,pixel$^{-1}$
and 0.73\AA\ (0.65\AA), respectively, which correspond to a \hb\ (\ha) FWHM velocity
resolution of about 34 (24) km\,s$^{-1}$. Each combined spectrum has a total exposure time  of 3600s-7200s. 
We reduced and calibrated the data using standard \textrm{IRAF} subroutines by following \citet{Fernandez2018}. Wavelength calibration was performed using CuNe+CuAr lamp arcs obtained immediately after the science exposures and have uncertainties of  $\lesssim$0.1\AA\ ($\sim$\,5 km\,s$^{-1}$). The spectra were corrected for atmospheric extinction and flux calibrated using spectrophotometric standard stars. Finally, 1D spectra were extracted using an optimal spatial aperture matching the spatial extent of the emission lines in the 2D spectra.


For the subsample of 13 galaxies, we used spectra obtained with the X-Shooter instrument at the Very Large Telescope (102.B-0942 and 106.215K; PI: D. Schaerer). The spectra were reduced using the ESO Reflex reduction pipeline \citep[version 2.11.5;][]{Freudling2013} to produce flux-calibrated spectra by \citet{Marques-Chaves2022}. For 5   of the  13 galaxies we also used an additional set of fully calibrated X-Shooter spectra presented and described by \citet{Guseva2020}. For these datasets, we only used the VIS arm, which has    a  spectral dispersion ($R\sim$\,9000) and total exposure times (3000s-6000s) that are comparable to those of our ISIS spectra.

\section{Emission-line modeling} \label{sec:methods}

We fitted emission line profiles of the galaxies following the multicomponent Gaussian method extensively described in \citet{Hogarth2020}. We used the fitting codes \textrm{FitELP} (Firpo et al. in prep.) and \textrm{LiMe}\footnote{\url{https://lime-stable.readthedocs.io/en/latest/}} \citep{Fernandez2023}, which are \textrm{Python} tools specifically designed to perform kinematic analysis of emission lines in high-resolution spectra and to fit emission line profiles using the Non-Linear Least-Squares Minimization and Curve-Fitting package \citep[LMFIT,][]{Newville2014}. For this work we deliberately restrict ourselves to the analysis of the following emission lines: \hb, [\oiii]$\lambda$5007, [\oi]$\lambda$6300, \ha, [\nii]$\lambda\lambda$6548,6584, and  [\sii]$\lambda\lambda$6717,6731.

The fitting process starts with the bright lines (\hb, [\oiii], and \ha) using a single Gaussian, and adds additional components based on the residuals and the reduced $\chi^2$ until the difference in the Akaike information criterion \citep[AIC][]{Akaike1974} between fits with $n$ and $n+1$ components remains minimal \citep[$\Delta$(AIC)$<$10;][]{Bosch2019}.
We fitted \ha\ and the [\nii] doublet simultaneously.
The peak velocity of each [\nii] component was initially linked to that of \ha\ and was allowed to vary a few pixels. The amplitude of [\nii]$\lambda$6548 was fixed to one-third of that in [\nii]$\lambda$6584, which is a free parameter. The central velocity and velocity dispersion of the \ha\ components were then copied to [\sii] and [\oi], which were fitted leaving free the amplitude of each component.
Finally, the intrinsic velocity dispersion ($\sigma$)
was obtained from the fitted value after subtracting in quadrature the instrumental and thermal velocity dispersion for each ion, as in \citet{Hogarth2020}.

In addition, we used \textsc{LiMe} to perform a non-parametric model-independent calculation of the $p$-th percentile velocities $v_{p}$ of emission lines (Figure~\ref{percentiles}).
We followed \citet{Liu2013} to derive the full width at zero intensity (FWZI) and the skewness or asymmetry parameter $A = ((v_{90}-v_{med})-(v_{med}-v_{10}))/ w_{80}$, with $v_{med}$ the median velocity and $w_{80}=v_{90}-v_{10}$ the velocity width enclosing 80\% of the total [\oiii] flux, respectively.

\section{Results and discussion} \label{sec:results}

\begin{figure}[t!]
\centering
\includegraphics[angle=0,width=0.436\textwidth]{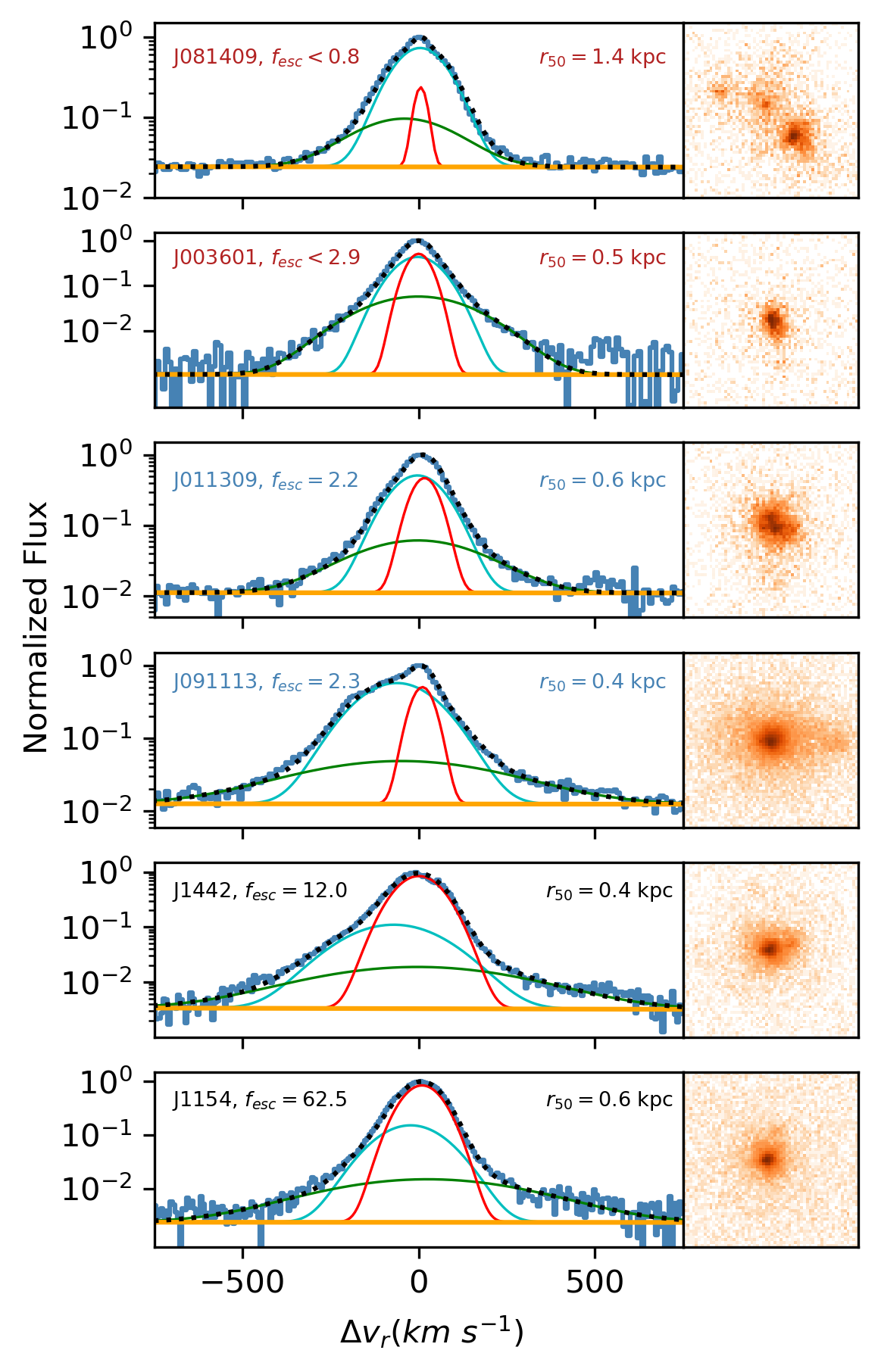}
\caption{Three-Gaussian model of [\oiii]5007 for a subsample observed with X-Shooter. Galaxies are shown in order of   increasing  LyC $f_{esc}$ percentage.  The  two upper panels show galaxies classified as NLCE,   the two middle panels show SLCEs, and the two bottom panels show  SLCEs. The right panels show \textit{HST-COS} NUV acquisition images of 1.4$"$ per side for each galaxy. The spectra are normalized to the peak flux and are  shown in log scale to highlight low-surface-brightness wings. \label{fig:f2}}
\end{figure}

Figures~\ref{fig:f1}-\ref{fig:f2} illustrate our Gaussian multicomponent line fitting. We show best-fit models for the \hb, \ha$+$[\nii], and [\oiii] observed profiles of a strong leaker (\objone) and the comparison of the [\oiii] modeling for strong and weak leakers, and for non-leakers.
We find that all the bright emission lines in our sample of galaxies cannot be modeled with a single Gaussian without leaving a strong residual that accounts for more than 50\% of the total line flux. Instead, they are best fitted as the superposition of three kinematic components.
Some LCEs need at least two resolved narrow components ($\sigma_{narrow}<$\,100\,\kms) to fit the emissions, while other LCEs are fitted with one narrow component and two broader components ($\sigma_{broad}>$\,100\,\kms).
The narrower component generally fits the main peak of the lines, while the broader component also fits  the wings of emission lines with $\sigma_{broad}$\,$\sim$\,120-300\,\kms. We find that the broad component is preferentially blueshifted from the systemic velocity derived from the peak of the lines ($\Delta v\sim|v_{broad} - v_{sys}|\sim$\,20-70\,\kms).
The complete kinematic analysis will be presented in Rodríguez-Henríquez et al. (in prep.).

\begin{figure*}[t!]
\begin{minipage}{1.0\textwidth}
\begin{center}
\includegraphics[angle=0,width=0.9\textwidth]{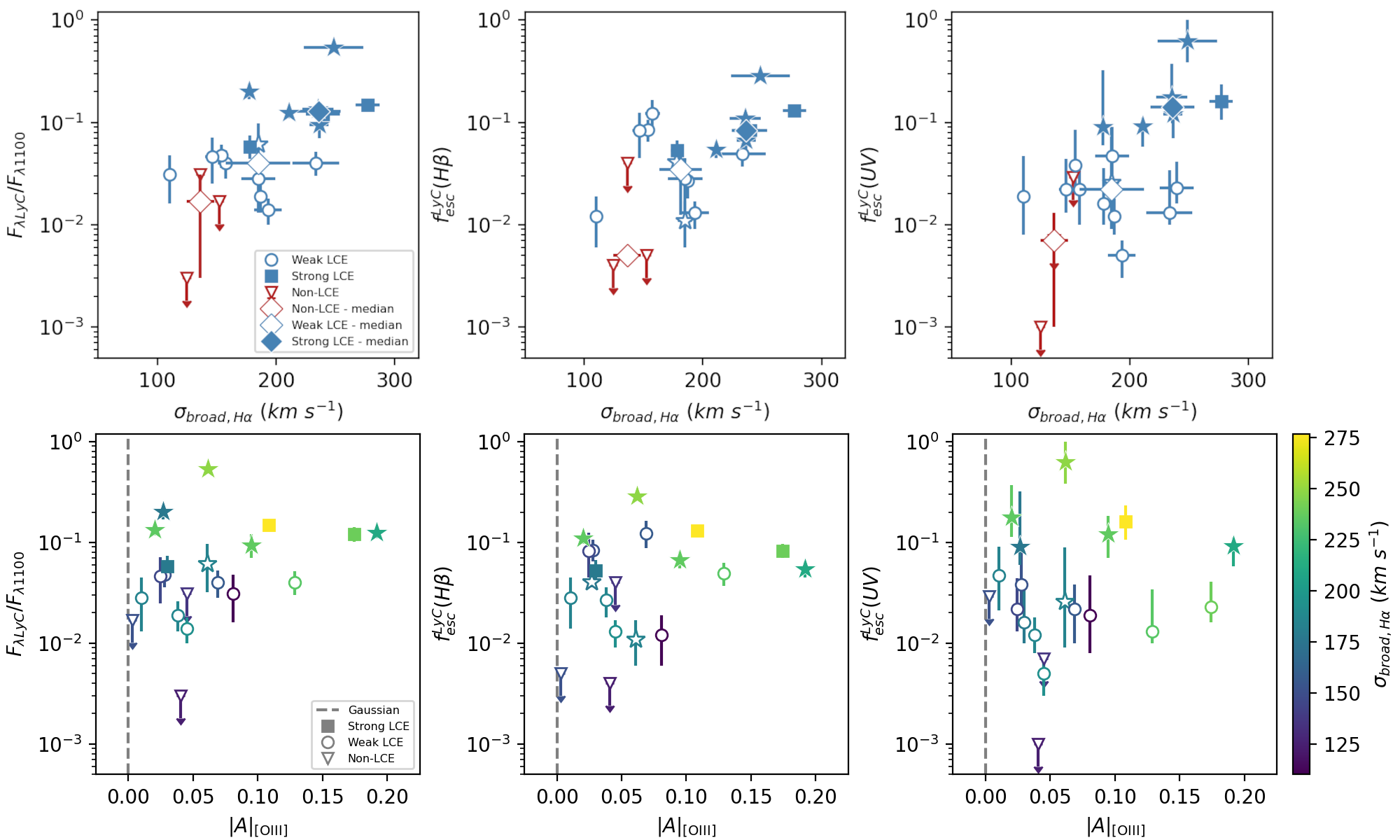}
\end{center}
\end{minipage}
\caption{Escape fraction traced by $F_{\lambda LyC}/F_{\lambda 1100}$ (left), $f_{esc}^{LyC}(H\beta)$ (middle), and $f_{esc}^{LyC}(UV)$ (right) vs. intrinsic velocity dispersion of the \ha\ broad component (top) and line asymmetry of [\oiii]\,5007 (bottom). The dotted lines ($A=0$) indicate a symmetric Gaussian profile. The stars identify the five galaxies from \citet{Izotov2016b,Izotov2018a}. \label{fig:f3}}
\end{figure*}

\subsection{Evidence of outflows of highly turbulent ionized gas}

Following \citet{Amorin2012b}, we interpret the narrower emissions as resulting from the virial motions of the photoionized nebulae surrounding the brightest unresolved young star clusters, which may follow the rotation of the gaseous disk \citep{Lofthouse2017, Bosch2019}. Instead, the broad emission, which largely exceeds the velocity dispersion ($\sim$60\,\kms) expected for rotating disks with M$_{\star}\lesssim$\,10$^{10}$\,M$_{\odot}$ \citep{Simons2015}, is ascribed to non-virial motions of high-velocity photoionized gas resulting from stellar feedback.
The lack of typical AGN spectral signatures \citep{Flury2022a} and the presence of
broad components in  Balmer lines and low- and high-ionization forbidden lines, strongly disfavor AGN broadening in this sample.
In addition, similar objects with spatially resolved spectra show broad emissions  extended to kiloparsec scales, which supports our  interpretation \citep[e.g.,][]{Bosch2019, Komarova2021}

Using a simple outflow model    \citep{Genzel2011}, the broader components are consistent with a spatially unresolved outflow with $v_{\rm out}=\Delta v + 2\sigma_{broad} \sim$\,200-600\,\kms.
They are likely the result of the strong winds of massive star clusters in the youngest and brightest unresolved star-forming clumps, and SNe feedback. Although we cannot resolve spatially the broad emission in our spectra, we find that galaxies with a more compact and often dominant UV central clump (Fig.~\ref{fig:f2}) have more complex line profiles and show broader wings, suggesting that the origin of the ionized outflows resides in the inner unresolved region of a few hundred parsecs or less.

\subsection{Properties of outflowing gas from line ratios}\label{sec:density}

In classic diagnostic diagrams, the excitation properties of the narrow and broad kinematic components of LCEs are consistent with line ratios driven by stellar photoionization (Fig.~\ref{bpt}). The narrow components show slightly higher [\oiii]/\hb\ and lower [\nii]/\ha\ than the broader components. However, for [\oi]/\ha\ and [\sii]/\ha\ we find that broad and narrow components tend to have lower ratios than normal galaxies, in line with the low  [\sii]/\ha\ found in a larger sample of LCEs using lower-resolution spectra \citep{Wang2021}.
This suggests the outflow traced by the broad component may tend to optically thin conditions \citep{Pellegrini2012} and have no significant contribution of shocks, in contrast to the higher [\sii]/\ha\ broad components seen in more massive galaxies \citep[e.g., luminous and ultra-luminous infrared galaxies, (U)LIRGs;][]{Arribas2014}.

In our sample, the broad component of [\sii] lines is significantly fainter and has larger uncertainty than for [\oiii] and \ha\ and \hb\  lines, so it should be taken with caution.
With this caveat in mind, we follow \citet{Fernandez2023} to derive [\sii] electron densities for the individual Gaussian components. 
We find relatively high densities ($n_{e} \sim$\,100 - 1000 cm$^{-3}$), with  values for the broad  
components that are comparable to or higher than the narrow components.

\begin{table*}[ht!]
\caption{Kendall-$\tau$ coefficients for LyC leakage and the H$\alpha$ and [\oiii] broad-line widths.}
\label{Table1}
\centering
\begin{tabular}{lcccccc}
\hline
 & \multicolumn{2}{c}{$F_{\lambda LyC}/F_{1100}$} & \multicolumn{2}{c}{$f_{esc}^{LyC}(H\beta)$} & \multicolumn{2}{c}{$f_{esc}^{LyC}(UV)$} \\
\cline{2-3} \cline{4-5} \cline{6-7}
 & $\tau$ & $p$ & $\tau$ & $p$ & $\tau$ & $p$ \\
\hline
$\sigma_{broad, H\alpha}$ & $0.46_{-0.12}^{+0.12}$ & $4.8\times 10^{-3}$ & $0.43_{-0.13}^{+0.12}$ & $8.6\times 10^{-3}$ & $0.43_{-0.12}^{+0.12}$ & $7.8\times 10^{-3}$ \\
$\sigma_{broad, [\mathrm{OIII}]}$ & $0.36_{-0.10}^{+0.09}$ & $2.5\times 10^{-2}$ & $0.25_{-0.14}^{+0.13}$ & $1.3\times 10^{-1}$ & $0.32_{-0.12}^{+0.10}$ & $5.2\times 10^{-2}$ \\
FWZI$_{[\mathrm{OIII}]}$ & $0.50_{-0.07}^{+0.09}$ & $2.1\times 10^{-3}$ & $0.20_{-0.11}^{+0.13}$ & $2.3\times 10^{-1}$ & $0.37_{-0.08}^{+0.08}$ & $2.3\times 10^{-2}$ \\
\hline
\end{tabular}
\medskip

\begin{minipage}{\textwidth} 
{\footnotesize Note: We list Kendall-$\tau$ coefficients for the three metrics of LyC leakage and the velocity dispersion of the  H$\alpha$ and [\oiii] broader component, and the [\oiii] full width at zero intensity (FWZI) (See Figs. \ref{fig:f3} and \ref{fig:fA.3}). Following \citet{Flury2022b}, a significant correlation is found if $p < 2.275 \times 10^{-2}$ and $|\tau| \geq 0.261$, while a weak correlation corresponds to values of $p < 1.587 \times 10^{-1}$ and $|\tau| \geq 0.162$.}
\end{minipage}
\end{table*}

\subsection{A correlation between outflow kinematics and LyC escape fraction}\label{sec:discussion1}

Revealing the nature of the broad emission in LCEs can provide new clues to how gas turbulence and stellar feedback affect \lya\ and LyC photon escape. Using a small sample of GPs, \citet{Amorin2012a} found broad emission with FWZI$\gtrsim$\,1000 \kms\ in recombination and forbidden nebular lines. More recent work suggests a causal connection between the complex kinematics and the high \lya\ escape fractions of GPs  \citep{Orlitova2018,Bosch2019, Hogarth2020} and lower-z analogs \citep{Herenz2016,Micheva2019,Komarova2021}.

In this work we explore the above tendency using higher-resolution data for a larger sample that contains confirmed LCEs. In Fig.~\ref{fig:f3} we present a correlation between $f^{LyC}_{esc}$ and the intrinsic velocity dispersion ($\sigma_{\rm broad}$) of the \ha\ broad component and line asymmetry $|A|$ (see \S3) of  [\oiii], respectively. We use the three metrics of $f^{LyC}_{esc}$ obtained and discussed by \citet{Flury2022a}: (i) the empirical ionizing--to--non-ionizing continuum ratio $F_{\lambda LyC}/F_{\lambda1100}$, (ii) using \hb\ flux to infer the intrinsic LyC from \textsc{starburst99} model spectra $f^{LyC}_{esc}$(\hb), and (iii) using SED fits to the far-UV spectra to infer the intrinsic LyC $f^{LyC}_{esc}$(UV).
In the upper panel of Fig.~\ref{fig:f3} the strongest leakers show a larger median velocity dispersion of $\sigma_{\rm broad}>$\,220\,\kms\ than the non-leakers, which show $\sigma_{\rm broad}<$\,150\,\kms. Using our previous definitions (\S4.1), this translates into median outflow velocities for strong leakers, weak leakers, and non-leakers of $v_{\rm out}=$\,485\,\kms, 333\,\kms, and 276\,\kms, respectively. As a comparison,
$v_{\rm out}\sim$\,327\,\kms\ have been reported for the strong LCE Sunburst Arc \citep{Mainali2022}.

The above correlation is robust against the metric considered for the LyC escape fraction and the chosen emission line (see also Fig. \ref{fig:fA.3}). This is shown in Table~\ref{Table1}, in which we quantify the significance of the above correlations using a  Kendall-$\tau$ test \citep{Akritas1996}. The above correlations are also evident if instead of $\sigma_{\rm broad}$ we use the velocity widths computed from our non-parametric analysis (Fig. \ref{percentiles}), such as the FWZI (Fig. \ref{fig:fA.3}) and  $w_{98}=v_{99}-v_{1}$  (i.e., the full width at the base of the line, excluding the noisy first and last 1\% of the total flux contained in the blue and red wings), for which we find $w_{98}=$\,640\,\kms\ and  $w_{98}=$\,484\,\kms\ for strong leakers and non-leakers, respectively.

In Fig.~\ref{fig:f3} (Bottom) we 
note a prevalence of asymmetric [\oiii] profiles ($|A|>0$), especially for LCEs of larger $\sigma_{\rm broad}$. Despite the large scatter, this relation appears evident when we compare the line profiles in Fig.\ref{O3-profiles}.
Quantitively, if we consider $v_{1}$, $v_{5}$, and $v_{10}$, which account for the velocity of the blueshifted emission line wing containing less than 1\%, 5\%, and 10\% of the total integrated flux, respectively, we find that SLCEs have median values that are  a factor of $\sim$\,1.5 larger than WLCES and NLCEs. However, no significant differences are found in the red wings ($v_{90}$, $v_{95}$, and $v_{99}$), suggesting that the trends in Fig.~\ref{fig:f3} are driven by the escape of LyC along the same line of sight of the outflow.

\subsection{Implications for LyC escape  mechanisms}\label{sec:implications}

In our sample of galaxies, LyC photons originate in young massive stars, and a fraction of them can escape through physical processes that are still not well understood.
In principle, LyC photons can either escape through density-bounded nebula or through low \hi\ column density holes or channels carved in the ISM  \citep[e.g.,][]{Zackrisson2013, Jaskot2013}. Observations of
LyC leakers have shown that the latter mechanism is largely at play
\citep[e.g.,][]{Reddy2016, Gazagnes2018, Saldana2022, Xu2023}. However, how these optically thin channels are created is unclear.
Simulations show that strong SNe feedback plays a key role in shaping the ISM and carving the holes through which LyC photons escape \citep[e.g.,][]{Trebitsch2017,Kimm2019}. In addition, intense stellar feedback injects strong turbulence in the ISM that can produce photoionized channels that favor LyC escape \citep{Kakiichi2021}.

We find a clear signature of turbulent ionized gas kinematics driven by stellar and SNe feedback in a sample of LCEs.
The width of the broad-line emission correlates with $f^{LyC}_{esc}$, providing new clues to the escaping mechanism.
Assuming that LyC photons in our LCEs emerge from the youngest starbursts ($\lesssim$5\,Myr), radiative feedback due to strong radiation pressure in the late stages of the evolution of massive stars can create an outflowing bubble of gas surrounding dense young star clusters. On this short timescale, the first SNe explode and produce mechanical energy and momentum, and cosmic ray feedback. This is further supported by the recent study of \citet{Bait2023}, who found evidence of SNe-driven non-thermal emission in the radio continuum observations of LzLCS galaxies, including 11 galaxies in our sample. In the youngest objects, however, the development of large-scale SNe-driven superwinds could be suppressed due to catastrophic cooling \citep{Jaskot2019,Komarova2021,Oey2023}. Thus, the relations we see in Fig.~\ref{fig:f3} may well be driven by radiatively driven outflows, which may carve dust-transparent ionized channels approaching density-bounded conditions \citep[e.g.,][]{Jaskot2013} that favor large escape fractions \citep{Ferrara2023}.

In this scenario the broad emission-line wings of LCEs could originate from the bulk outflow near the starburst \citep{Martin2015} and from 
dense pockets of highly ionized gas interacting
with a high-velocity wind fluid, possibly entrained near the base of a larger-scale wind
and close to the ionization source \citep{Hayes2023} or in the shell of the highly ionized bubble.
On longer timescales ($\gtrsim$5-10\,Myr), SNe feedback should take over when the LyC production from massive stars declines rapidly \citep[e.g.,][]{Naidu2022}. Outflows may have time enough to break the ionizing bubbles and produce larger cavities and holes in the ISM. Results from current models and simulations suggest that the two feedback modes could eventually operate sequentially \citep[e.g.,][]{Kakiichi2021,Katz2023}.

Observationally, a two-stage starburst has been proposed to explain the ionizing photon escape in some galaxies \citep{Micheva2018}. In this scenario, shocks induced by recent SNe explosions from an initial starburst episode with ages $>$\,10\,Myr may have   enough time to clear out the surrounding ISM allowing the LyC photon production of a second younger starburst ($<$5\, Myr) to have an easier escaping path \citep[e.g.,][]{Enders2023}.

Triggered star formation in the denser shells of cold gas shaped by the mechanical feedback from a previous starburst episode has been reported in galactic and extragalactic nebulae \citep[e.g.,][]{Egorov2023}. The younger massive stars emerging from these regions can produce large amounts of ionizing photons that may escape from the nebula, as predicted by simulations \citep{Ma2020}. This scenario could be interesting to explain the observed scatter in the different relations between $f^{LyC}_{esc}$ and global properties \citep[e.g.,][]{Flury2022b}. Nevertheless, it is difficult to probe because the ongoing starburst generally dominates the observed nebular emission, and resolving the short timespan between two recent bursts from optical spectral synthesis is challenging, even with very high S/N spectra \citep{Amorin2012a,Fernandez2022}. Future spatially resolved UV to near-IR spectroscopy will help to probe this scenario.

Finally, geometric effects have an important role in the connection between gas kinematics and the escape of ionizing photons \citep[e.g.,][]{Zastrow2011,Bassett2019,Ramambason2020,Carr2021}.
LyC escape could be highly anisotropic and viewing-geometry dependent \citep{Kim2023}. The asymmetry of line profiles with a prevalence of blueshifted emission in our LCEs could indicate anisotropic leakage through favorable pencil-beam sightlines that are less affected by dust obscuration \citep{Martin2015}. We note that the stronger LCEs tend to have the most compact and least attenuated starbursts (i.e., very high SFR surface densities and $\beta_{UV}$ slopes) \citep[see also][]{Flury2022b,Chisholm2022}.
This discussion will be expanded in a future study via the detailed comparison of the UV and optical line profiles.

\section{Conclusions and outlook}

In this work we present new high-resolution optical spectra of 6 strong ($f^{LyC}_{esc}$\,$\sim$\,5-63\%) and  11 weak LyC emitters ($f^{LyC}_{esc}$\,$\sim$\,2-5\%), and 3 galaxies without significant LyC leakage ($f^{LyC}_{esc}$\,$<$\,2\%). Using this data we performed a first kinematic analysis of resolved emission-line profiles using multicomponent Gaussian fitting.

Our results demonstrate the ubiquitous presence of broad emission-line wings ($\sigma_{\rm broad}\sim$\,100-300 \kms), often blueshifted ($\Delta v_r\sim$\,20-70\,\kms), tracing high-velocity emission-line components (FWZI$\gtrsim$\,750 \kms) that underlie narrower components ($\sigma_{\rm narrow}$\,$\sim$\,40-100\,\kms). The narrow emission traces H{\sc ii} regions, which follow the disk kinematics and are associated with  young bright star-forming regions responsible for the production of the ionizing photons. The broad emission is instead  interpreted as a turbulent photoionized gas tracing an unresolved outflow, likely driven by starburst feedback, namely strong radiation pressure, winds of young massive stars, and SNe.

We find a significant correlation between the intrinsic velocity dispersion and maximum line-width velocities of galaxies and their LyC escape fraction. Thus, strong LyC leakers show stronger broad components with larger line widths and a prevalence of larger asymmetries than weak leakers  and non-leakers. This kinematic complexity of strong leakers contrasts with their otherwise simpler UV morphology from HST/COS imaging.

Although SNe-driven feedback should play a role in the entire galaxy sample \citep{Bait2023}, we speculate that in the stronger LCEs, the broad emission primarily emerges from the radiation pressure and strong winds of massive stars in the highly pressurized environment of extremely young and unresolved ($<$\,250 pc) star-forming clumps that dominate the UV luminosity budget of these galaxies.

Overall, our results strongly suggest that the physical mechanisms driving the observed kinematic complexity play a significant role in the escape of ionizing photons in galaxies. This adds new observational support to predictions of models and simulations \citep[e.g.,][]{Trebitsch2017, Kimm2019, Kakiichi2021}, in which ongoing starbursts and their related radiative and mechanical feedback produce gas turbulence and outflows that are key in clearing channels through which ionizing radiation escapes into the intergalactic medium.

Future work will increase the number of galaxies with LyC observations at low and high redshifts, for which medium- or high-resolution spectra with ground-based spectrographs could extend this relation. Different fitting techniques and physically motivated models will also be explored \citep{Komarova2021,Flury2023}. In addition, spatially resolved high-resolution spectra are needed to identify the source of the outflow and the ionizing photons in LCEs.
We find that detection of broad-line components in GPs 
could be indicative of LyC leakage. Features like these can be probed in reionization galaxies with the JWST/NIRSpec \citep[e.g.,][]{Carniani2023,Xu-Ouchi2023}. Finally, future HST/COS high-resolution UV spectra are needed to better constrain the outflows seen in absorption and to compare \lya\ and optical emission-line profiles, whereas HST optical narrowband imaging may provide additional hints for constraining the geometry of the emitting regions.

\begin{acknowledgements}
We thank the anonymous referee for the prompt and helpful report.  
RA acknowledges the support of ANID FONDECYT Regular Grant 1202007 and DIDULS/ULS PTE2053851. VF and DM acknowledge support from ANID FONDECYT Postdoctoral grant 3200473 and ANID Scolarship Program 2019-21191543, respectively. 
JMV acknowledges financial support from the State Agency for
Research of the Spanish MCIU through ‘Center of Excellence Severo Ochoa’
award to the IAA-CSIC (SEV-2017-0709) and CEX2021-001131-S and PID2022-136598NB-C32 funded by MCIN/AEI/10.13039/501100011033, and from project PID2019-107408GB-C44. 
N.G. and Y.I. acknowledge support from the National Academy of Sciences of Ukraine (Project No. 0121U109612). ASL acknowledges support from Knut and Alice Wallenberg Foundation.

\end{acknowledgements}

%

\begin{thebibliography}{}

\bibitem[Akaike(1974)]{Akaike1974} Akaike, H.\ 1974, IEEE Transactions on Automatic Control, 19, 716
\bibitem[Akritas \& Siebert(1996)]{Akritas1996} Akritas, M.~G. \& Siebert, J.\ 1996, \mnras, 278, 919 
\bibitem[Amor{\'\i}n et al.(2010)]{Amorin2010} Amor{\'\i}n, R.~O., P{\'e}rez-Montero, E., \& V{\'\i}lchez, J.~M.\ 2010, \apjl, 715, L128
\bibitem[Amor{\'\i}n et al.(2012a)]{Amorin2012a} Amor{\'\i}n, R., P{\'e}rez-Montero, E., V{\'\i}lchez, J.~M., et al.\ 2012, \apj, 749, 185
\bibitem[Amor{\'\i}n et al.(2012b)]{Amorin2012b} Amor{\'\i}n, R., V{\'\i}lchez, J.~M., H{\"a}gele, G.~F., et al.\ 2012, \apjl, 754, L22
\bibitem[Arribas et al.(2014)]{Arribas2014} Arribas, S., Colina, L., Bellocchi, E., et al.\ 2014, \aap, 568, A14
\bibitem[Bait et al.(2023)]{Bait2023} Bait, O., Borthakur, S., Schaerer, D., et al.\ 2023, arXiv:2310.18817 
\bibitem[Barrow et al.(2020)]{Barrow2020} Barrow, K.~S.~S., Robertson, B.~E., Ellis, R.~S., et al.\ 2020, \apjl, 902, L39 
\bibitem[Bassett et al.(2019)]{Bassett2019} Bassett, R., Ryan-Weber, E.~V., Cooke, J., et al.\ 2019, \mnras, 483, 5223 
\bibitem[Bosch et al.(2019)]{Bosch2019} Bosch, G., H{\"a}gele, G.~F., Amor{\'\i}n, R., et al.\ 2019, \mnras, 489, 1787
\bibitem[Cardamone et al.(2009)]{Cardamone2009} Cardamone, C., Schawinski, K., Sarzi, M., et al.\ 2009, \mnras, 399, 1191
\bibitem[Carniani et al.(2023)]{Carniani2023} Carniani, S., Venturi, G., Parlanti, E., et al.\ 2023, arXiv:2306.11801 
\bibitem[Carr et al.(2021)]{Carr2021} Carr, C., Scarlata, C., Henry, A., et al.\ 2021, \apj, 906, 104 
\bibitem[Chisholm et al.(2017)]{Chisholm2017} Chisholm, J., Orlitov{\'a}, I., Schaerer, D., et al.\ 2017, \aap, 605, A67
\bibitem[Chisholm et al.(2022)]{Chisholm2022} Chisholm, J., Saldana-Lopez, A., Flury, S., et al.\ 2022, \mnras 
\bibitem[Egorov et al.(2023)]{Egorov2023} Egorov, O.~V., Kreckel, K., Glover, S.~C.~O., et al.\ 2023, \aap, 678, A153 
\bibitem[Enders et al.(2023)]{Enders2023} Enders, A.~U., Bomans, D.~J., \& Wittje, A.\ 2023, \aap, 672, A11 
\bibitem[Fern{\'a}ndez et al.(2018)]{Fernandez2018} Fern{\'a}ndez, V., Terlevich, E., D{\'\i}az, A.~I., et al.\ 2018, \mnras, 478, 5301 
\bibitem[Fern{\'a}ndez et al.(2022)]{Fernandez2022} Fern{\'a}ndez, V., Amor{\'\i}n, R., P{\'e}rez-Montero, E., et al.\ 2022, \mnras, 511, 2515 
\bibitem[Fern{\'a}ndez et al.(2023)]{Fernandez2023} Fern{\'a}ndez, V., Amor{\'\i}n, R., Sanchez-Janssen, R., et al.\ 2023, \mnras, 520, 3576 
\bibitem[\protect\citeauthoryear{Ferrara}{2023}]{Ferrara2023} Ferrara A., 2023, arXiv, arXiv:2310.12197 
\bibitem[Flury et al.(2022a)]{Flury2022a} Flury, S.~R., Jaskot, A.~E., Ferguson, H.~C., et al.\ 2022a, \apjs, 260, 1 
\bibitem[Flury et al.(2022b)]{Flury2022b} Flury, S.~R., Jaskot, A.~E., Ferguson, H.~C., et al.\ 2022b, \apj, 930, 126 
\bibitem[Flury et al.(2023)]{Flury2023} Flury, S.~R., Moran, E.~C., \& Eleazer, M.\ 2023, \mnras 
\bibitem[Freudling et al.(2013)]{Freudling2013} Freudling, W., Romaniello, M., Bramich, D.~M., et al.\ 2013, \aap, 559, A96 
\bibitem[Gazagnes et al.(2018)]{Gazagnes2018} Gazagnes, S., Chisholm, J., Schaerer, D., et al.\ 2018, \aap, 616, A29 
\bibitem[Gazagnes et al.(2020)]{Gazagnes2020} Gazagnes, S., Chisholm, J., Schaerer, D., et al.\ 2020, \aap, 639, A85 
\bibitem[Genzel et al.(2011)]{Genzel2011} Genzel, R., Newman, S., Jones, T., et al.\ 2011, \apj, 733, 101
\bibitem[Guseva et al.(2020)]{Guseva2020} Guseva, N.~G., Izotov, Y.~I., Schaerer, D., et al.\ 2020, \mnras, 497, 4293 
\bibitem[Heckman et al.(2011)]{Heckman2011} Heckman, T.~M., Borthakur, S., Overzier, R., et al.\ 2011, \apj, 730, 5
\bibitem[Herenz et al.(2016)]{Herenz2016} Herenz, E.~C., Gruyters, P., Orlitova, I., et al.\ 2016, \aap, 587, A78
\bibitem[Hayes(2023)]{Hayes2023} Hayes, M.~J.\ 2023, \mnras, 519, L26 
\bibitem[Hogarth et al.(2020)]{Hogarth2020} Hogarth, L., Amor{\'\i}n, R., V{\'\i}lchez, J.~M., et al.\ 2020, \mnras, 494, 3541 
\bibitem[Izotov et al.(2016a)]{Izotov2016a} Izotov, Y.~I., Orlitov{\'a}, I., Schaerer, D., et al.\ 2016, \nat, 529, 178
\bibitem[Izotov et al.(2016b)]{Izotov2016b} Izotov, Y.~I., Schaerer, D., Thuan, T.~X., et al.\ 2016, \mnras, 461, 3683
\bibitem[Izotov et al.(2018a)]{Izotov2018a} Izotov, Y.~I., Schaerer, D., Worseck, G., et al.\ 2018, \mnras, 474, 4514
\bibitem[Izotov et al.(2018b)]{Izotov2018b} Izotov, Y.~I., Worseck, G., Schaerer, D., et al.\ 2018, \mnras, 478, 4851
\bibitem[Jaskot \& Oey(2013)]{Jaskot2013} Jaskot, A.~E., \& Oey, M.~S.\ 2013, \apj, 766, 91
\bibitem[Jaskot et al.(2017)]{Jaskot2017} Jaskot, A.~E., Oey, M.~S., Scarlata, C., et al.\ 2017, \apjl, 851, L9
\bibitem[Jaskot et al.(2019)]{Jaskot2019} Jaskot, A.~E., Dowd, T., Oey, M.~S., et al.\ 2019, \apj, 885, 96 
\bibitem[Kakiichi \& Gronke(2021)]{Kakiichi2021} Kakiichi, K. \& Gronke, M.\ 2021, \apj, 908, 30 
\bibitem[Katz et al.(2023)]{Katz2023} Katz, H., Saxena, A., Rosdahl, J., et al.\ 2023, \mnras, 518, 270 
\bibitem[Kim et al.(2020)]{Kim2020} Kim, K., Malhotra, S., Rhoads, J.~E., et al.\ 2020, \apj, 893, 134 
\bibitem[Kim et al.(2023)]{Kim2023} Kim, K.~J., Bayliss, M.~B., Rigby, J.~R., et al.\ 2023, \apjl, 955, L17 
\bibitem[Kimm \& Cen(2014)]{KimmCen2014} Kimm, T. \& Cen, R.\ 2014, \apj, 788, 121
\bibitem[Kimm et al.(2019)]{Kimm2019} Kimm, T., Blaizot, J., Garel, T., et al.\ 2019, \mnras, 486, 2215
\bibitem[Komarova et al.(2021)]{Komarova2021} Komarova, L., Oey, M.~S., Krumholz, M.~R., et al.\ 2021, \apjl, 920, L46.
\bibitem[Liu et al.(2013)]{Liu2013} Liu, G., Zakamska, N.~L., Greene, J.~E., et al.\ 2013, \mnras, 436, 2576.
\bibitem[Lofthouse et al.(2017)]{Lofthouse2017} Lofthouse, E.~K., Houghton, R.~C.~W., \& Kaviraj, S.\ 2017, \mnras, 471, 2311
\bibitem[Llerena et al.(2023)]{Llerena2023} Llerena, M., Amor{\'\i}n, R., Pentericci, L., et al.\ 2023, \aap, 676, A53 
\bibitem[Ma et al.(2016)]{Ma2016} Ma, X., Hopkins, P.~F., Kasen, D., et al.\ 2016, \mnras, 459, 3614
\bibitem[Ma et al.(2020)]{Ma2020} Ma, X., Quataert, E., Wetzel, A., et al.\ 2020, \mnras, 498, 2001 
\bibitem[Mainali et al.(2022)]{Mainali2022} Mainali, R., Rigby, J.~R., Chisholm, J., et al.\ 2022, \apj, 940, 160 
\bibitem[Marques-Chaves et al.(2022a)]{Marques-Chaves2022a} Marques-Chaves, R., Schaerer, D., {\'A}lvarez-M{\'a}rquez, J., et al.\ 2022a, \mnras, 517, 2972 
\bibitem[Marques-Chaves et al.(2022b)]{Marques-Chaves2022} Marques-Chaves, R., Schaerer, D., Amor{\'\i}n, R.~O., et al.\ 2022b, \aap, 663, L1 
\bibitem[Martin et al.(2015)]{Martin2015} Martin, C.~L., Dijkstra, M., Henry, A., et al.\ 2015, \apj, 803, 6 
\bibitem[Matthee et al.(2021)]{Matthee2021} Matthee J., Sobral D., Hayes M., et al.\ 2021, \mnras, 505, 1382.
\bibitem[Micheva et al.(2018)]{Micheva2018} Micheva, G., Oey, M.~S., Keenan, R.~P., et al.\ 2018, \apj, 867, 2
\bibitem[Micheva et al.(2019)]{Micheva2019} Micheva, G., Christian Herenz, E., Roth, M.~M., et al.\ 2019, \aap, 623, A145
\bibitem[Naidu et al.(2022)]{Naidu2022} Naidu, R.~P., Matthee, J., Oesch, P.~A., et al.\ 2022, \mnras, 510, 4582 
\bibitem[Nakajima et al.(2013)]{Nakajima2013} Nakajima, K., Ouchi, M., Shimasaku, K., et al.\ 2013, \apj, 769, 3
\bibitem[Nelson et al.(2019)]{Nelson2019} Nelson, D., Pillepich, A., Springel, V., et al.\ 2019, \mnras, 490, 3234 
\bibitem[Newville et al.(2014)]{Newville2014} Newville, M., Stensitzki, T., Allen, D.~B., et al.\ 2014, Zenodo
\bibitem[Orlitov{\'a} et al.(2018)]{Orlitova2018} Orlitov{\'a}, I., Verhamme, A., Henry, A., et al.\ 2018, \aap, 616, A60
\bibitem[Oey et al.(2023)]{Oey2023} Oey, M.~S., Sawant, A.~N., Danehkar, A., et al.\ 2023, \apjl, 958, L10 
\bibitem[Pellegrini et al.(2012)]{Pellegrini2012} Pellegrini, E.~W., Oey, M.~S., Winkler, P.~F., et al.\ 2012, \apj, 755, 40 
\bibitem[Ramambason et al.(2020)]{Ramambason2020} Ramambason, L., Schaerer, D., Stasi{\'n}ska, G., et al.\ 2020, \aap, 644, A21 
\bibitem[Reddy et al.(2016)]{Reddy2016} Reddy, N.~A., Steidel, C.~C., Pettini, M., et al.\ 2016, \apj, 828, 108 
\bibitem[Rosdahl et al.(2018)]{Rosdahl2018} Rosdahl, J., Katz, H., Blaizot, J., et al.\ 2018, \mnras, 479, 994 
\bibitem[Saldana-Lopez et al.(2022)]{Saldana2022} Saldana-Lopez, A., Schaerer, D., Chisholm, J., et al.\ 2022, \aap, 663, A59
\bibitem[Simons et al.(2015)]{Simons2015} Simons, R.~C., Kassin, S.~A., Weiner, B.~J., et al.\ 2015, \mnras, 452, 986
\bibitem[Trebitsch et al.(2017)]{Trebitsch2017} Trebitsch, M., Blaizot, J., Rosdahl, J., et al.\ 2017, \mnras, 470, 224 
\bibitem[Vanzella et al.(2022)]{Vanzella2022} Vanzella, E., Castellano, M., Bergamini, P., et al.\ 2022, \aap, 659, A2 
\bibitem[Wang et al.(2021)]{Wang2021} Wang, B., Heckman, T.~M., Amor{\'\i}n, R., et al.\ 2021, \apj, 916, 3 
\bibitem[Wise \& Cen(2009)]{WiseCen2009} Wise, J.~H., \& Cen, R.\ 2009, \apj, 693, 984
\bibitem[Whittle (1985)]{Whittle1985} Whittle, M.\ 1985, \mnras, 213, 1.
\bibitem[Xu et al.(2023)]{Xu2023} Xu, X., Heckman, T., Henry, A., et al.\ 2023, \apj, 948, 28 
\bibitem[Xu et al.(2023)]{Xu-Ouchi2023} Xu, Y., Ouchi, M., Nakajima, K., et al.\ 2023, arXiv:2310.06614 
\bibitem[Zackrisson et al.(2013)]{Zackrisson2013} Zackrisson, E., Inoue, A.~K., \& Jensen, H.\ 2013, \apj, 777, 39 
\bibitem[Zastrow et al.(2011)]{Zastrow2011} Zastrow, J., Oey, M.~S., Veilleux, S., et al.\ 2011, \apjl, 741, L17 
\bibitem[Zastrow et al.(2013)]{Zastrow2013} Zastrow, J., Oey, M.~S., Veilleux, S., et al.\ 2013, \apj, 779, 76 

\end{thebibliography}
%


\begin{appendix} 
\section{Additional figures}

\begin{figure*}
\centering
\includegraphics[width=10.9cm]{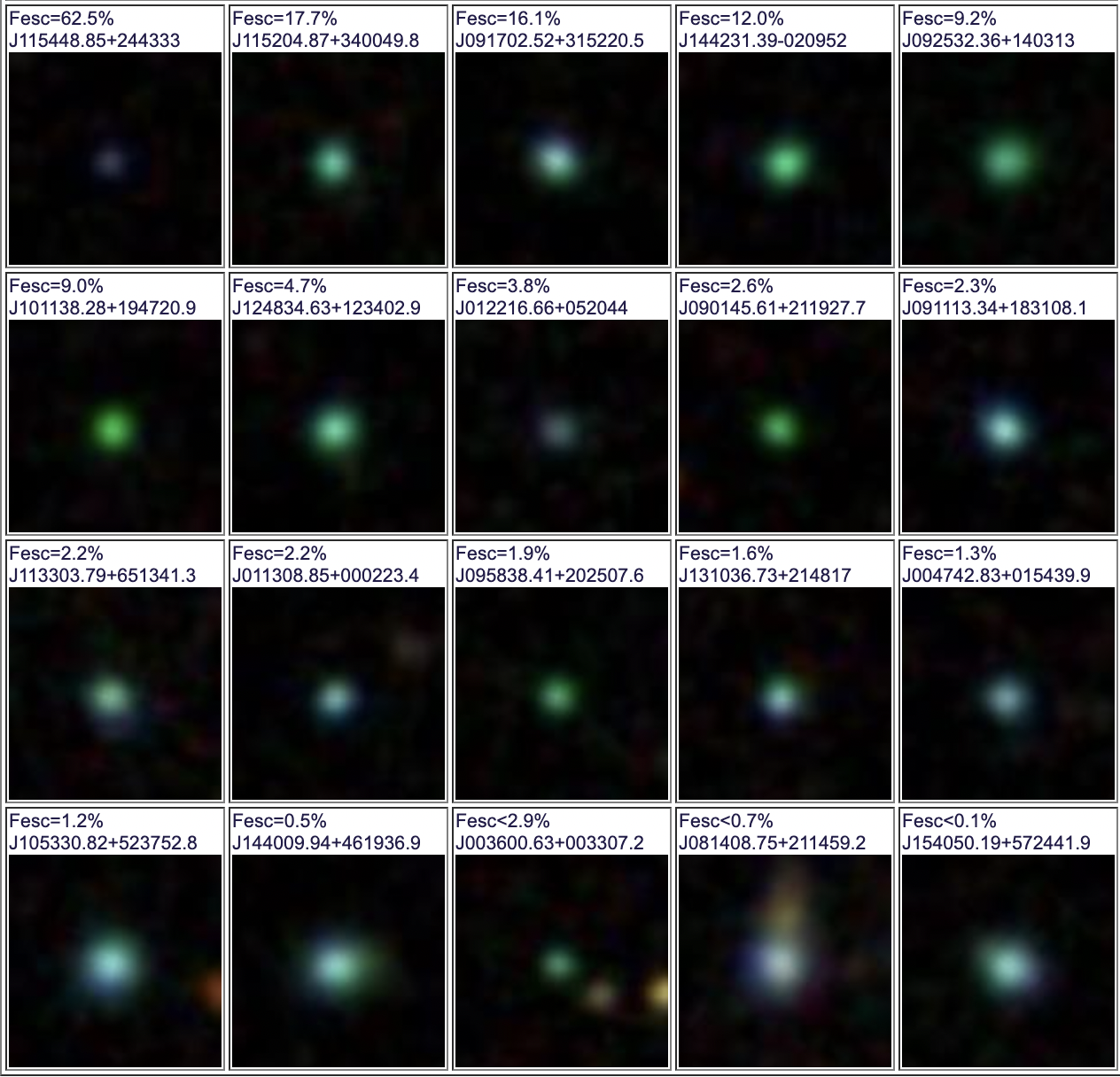}
\caption{Sample of galaxies. SDSS-DR16 \textit{ugriz} color cutouts are displayed. Galaxies are shown from top left to bottom right in order of  decreasing  $f^{LyC}_{esc}$  percentage \citep{Flury2022a}}
\label{sample}
\end{figure*}

\begin{figure}
\centering
\includegraphics[width=8.9cm]{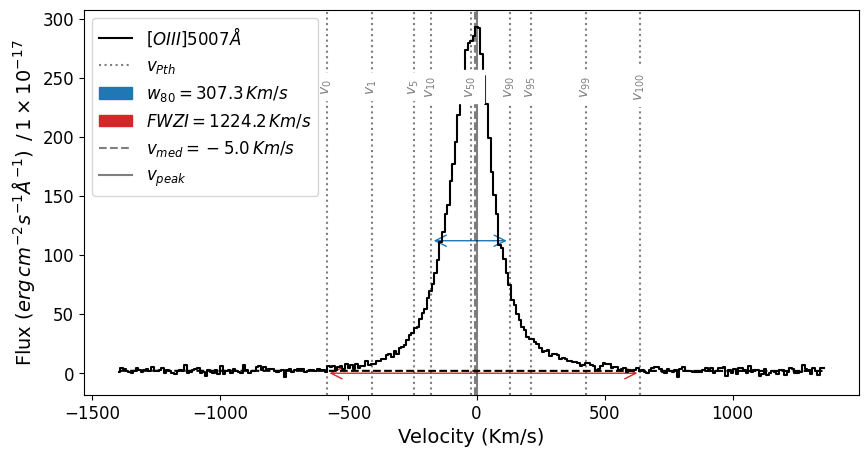}
\caption{Illustration of the inter-percentile analysis performed by LiMe \citep[][fully documented in \url{https://lime-stable.readthedocs.io/en/latest/}]{Fernandez2023}. Shown are the [\oiii]\,5007 line profile of the LCE \mbox{$J$115204+340049}. The velocity percentiles $v_p$ enclosing $p\%$ of the total line integrated flux are indicated with vertical dotted lines. The red arrow line illustrates the full width at zero intensity (FWZI) computed as $v_{100}-v_{0}$. The inset also shows values for the derived relevant quantities, including $w_{80}$ (blue arrow) and the median ($v_{med}$) and peak velocity ($v_{peak}$), the latter assumed to be the systemic velocity (see Section~2). }
\label{percentiles}
\end{figure}

\begin{figure}
\centering
\includegraphics[width=8.cm]{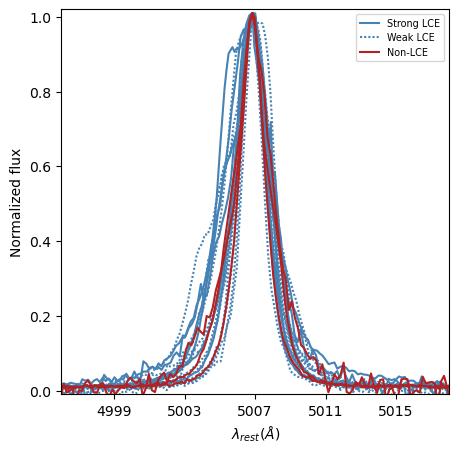}
\caption{Superposition of normalized, continuum-subtracted, rest-frame  [\oiii]\,5007 line profiles for the galaxy sample. This illustrates the differences in asymmetry and the extent of emission line wings for SLCEs and WLCEs (blue solid and dotted line, respectively) and  NLCEs (red). }
\label{O3-profiles}
\end{figure}

\begin{figure*}
\centering
\includegraphics[width=12.4cm]{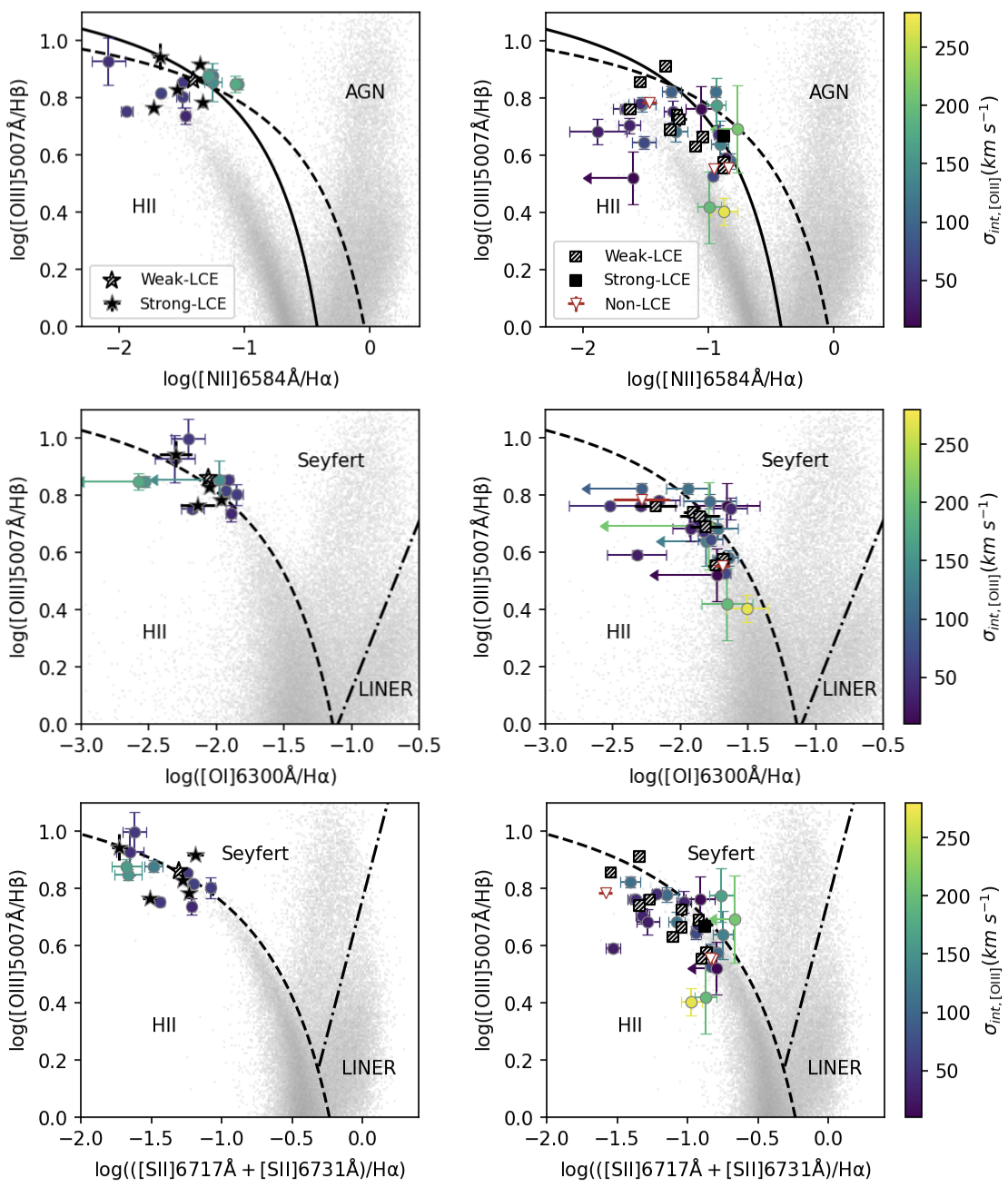}
\caption{Classic diagnostic diagrams. Shown are gas excitation [\oiii]/\hb\  vs. $\log$([\nii]/\ha) (\textit{top}), vs. $\log$([\oi]/\ha) (\textit{middle}), and vs. $\log$([\sii]/\ha) (\textit{bottom}). The left and right panels show the \citep{Izotov2016a,Izotov2016b,Izotov2018a} and LzLCS  \citep{Wang2021,Flury2022a} samples (Section~2), respectively. Stars, squares, and open triangles represent integrated values. Circles indicate {individual} Gaussian components;   the color bar gives their intrinsic velocity dispersion.}
\label{bpt}
\end{figure*}

\begin{figure*}[t!]
\centering
\includegraphics[angle=0,width=0.9\textwidth]{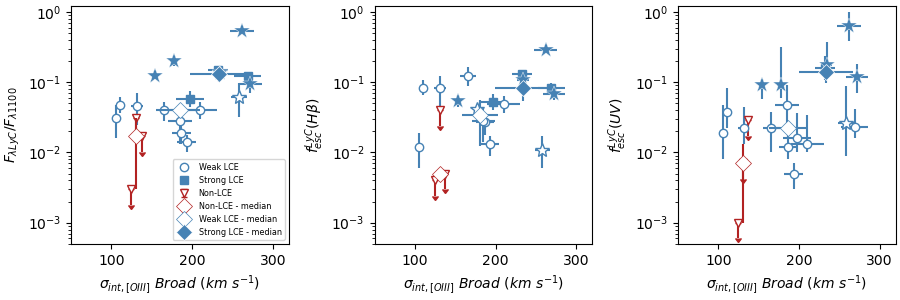} \includegraphics[angle=0,width=0.9\textwidth]{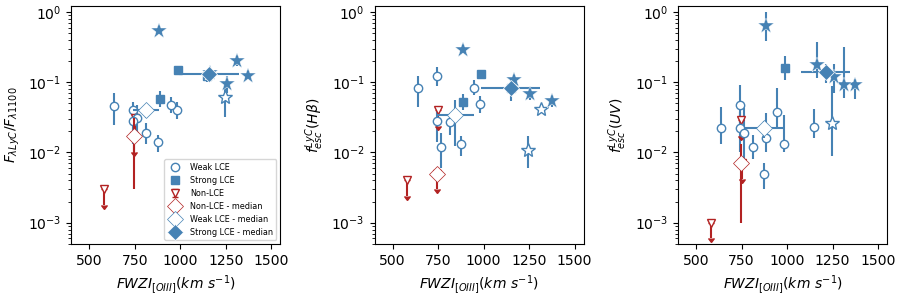}
\caption{Escape fraction traced by $F_{\lambda LyC}/F_{\lambda 1100}$ (left), $f_{esc}^{LyC}(H\beta)$ (middle), and $f_{esc}^{LyC}(UV)$ (right) vs. intrinsic velocity dispersion of the broad component (top) and full-width at zero intensity of the [\oiii]\,5007 lines (bottom). The symbols are as in Fig.~\ref{fig:f3}. \label{fig:fA.3}}
\end{figure*}

\end{appendix}
\end{document}